# Value of Agreement in Decision Analysis: Concept, Measures and Application

Tom Pape*

**Abstract:** In multi-criteria decision analysis workshops, participants often appraise the options individually before discussing the scoring as a group. The individual appraisals lead to score ranges within which the group then seeks the necessary agreement to identify their preferred option. Preference programming enables some options to be identified as dominated even before the group agrees on a precise scoring for them.

Workshop participants usually face time pressure to make a decision. Decision support can be provided by flagging options for which further agreement on their scores seems particularly valuable. By valuable, we mean the opportunity to identify other options as dominated (using preference programming) without having their precise scores agreed beforehand. The present paper quantifies this Value of Agreement and extends the concept to portfolio decision analysis and criterion weights. The new concept is validated through a case study in recruitment.

---

**1 Introduction**

Multi-criteria decision analysis (MCDA) often assesses options with an additive independent value function (Keeney & Raiffa 1976). The performance of each option $i \in I$ is valued on each criterion $j \in J$ along a 0–100 scale to obtain the value scores $v_{ij}$ (see Table 1 for symbols). While the value score for tangible criteria can often be derived from well-defined, marginal value functions, which map attribute levels to value scores (e.g. frequency of a service, horsepower of a machine, traffic noise), many intangible criteria require highly subjective judgements to assign value scores to options (e.g. impact of constructions on scenery, level of expertise of organisation, comfort of vehicle) (Keeney 1982). The criteria are weighted against each other, leading to the relative importance judgements $w_j$ with $\sum_j w_j = 1$. The decision model recommends the option with the greatest score $V_i = \sum_j w_j v_{ij}$.

Real-world problems are rarely that simplistic, and extensions to the model are therefore required. The present paper addresses MCDA problems in which the score preferences are initially incomplete; i.e. the criterion scores can take any number in the range $[\underline{v}_{ij}, \bar{v}_{ij}]$. The options' value scores hence span from $\underline{V}_i = \sum_j w_j \underline{v}_{ij}$ to $\bar{V}_i = \sum_j w_j \bar{v}_{ij}$. If there is one option that dominates all others, it is called the robust option (Roy 1998).

Such incomplete preference information (Hazen 1986, Kirkwood & Sarin 1985, Sarin 1977) is often encountered by decision-making groups appraising options against intangible selection criteria. In multi-criteria group decision making, participants frequently appraise the options individually before seeking agreement on the scores. The individual precise criterion scores $v_{ijk}$ by participants $k \in K$ lead to criterion score ranges from $\underline{v}_{ij} = \min_k\{v_{ijk}\}$ to $\bar{v}_{ij} = \max_k\{v_{ijk}\}$, within which the group looks for the necessary agreement to identify the group's robust option. 'Aggregation' and 'consensus' are the two principle methods to reach this necessary agreement. Using the aggregation method, an influence weight is assigned to each participant (e.g. Dias & Sarabando 2012, Kim & Ahn 1999, Salo 1995). A precise group score for all decision options can be automatically calculated by weighting the individual scores, which immediately clarifies the group's preferred option according to the participants' influence. Using the consensus method, the group engages in a discussion from which the necessary consensus on the scores needs to emerge with mutual agreement (e.g. Dias & Climaco 2005, Matsatsinis & Samaras 2001, Mustajoki et al. 2007, Phillips 2011, Shakhsi-Niaei et al. 2011, Vilkkumaa et al. 2014). The aggregation method stresses the positional power of each decision maker, while the consensus method emphasises mutual learning from each other's insights in the decision problem. When using the consensus method, time pressure may make it challenging for the group to come to a full agreement on all scores (Kim & Ahn 1997, Weber 1987). In this paper, we develop a new concept, which helps the consensus method in multi-criteria group decision-making to be more time-effective.

---

* Email: t.pape@surrey.ac.uk.

When thoroughly reviewing the individual appraisals $V_{ik} = \sum_j w_j v_{ijk}$ and the resulting score ranges from $\underline{V_i} = \min_k \{V_{ik}\}$ to $\bar{V}_i = \max_k \{V_{ik}\}$, a skilled facilitator might be able to roughly guess which options are definitely dominated, which options are likely to be dominated after further agreement on some other options is sought, which ones have a totally unclear fate and which ones have a good chance of becoming the robust one. The present paper attempts to replace this intuitive guessing with a quantitative concept we call Value of Agreement. Options with a high Value of Agreement are more attractive to a group discussion as knowing their agreed value score may allow eliminating other options as dominated without the need to seek agreement on their scores beforehand; thus, finding the robust option will require less effort. Altogether, the Value of Agreement estimates the impact of eliciting additional preference information on the time requirements of a decision analysis workshop.

Designing workshops to eliminate options quickly has hitherto been addressed only by very few publications. Hämäläinen & Pöyhönen (1996) and Salo & Hämäläinen (1995) analysed problems where only the weight information was incomplete. They suggested that the group should first seek agreement on the criteria with large weight ranges. Mustajoki & Hämäläinen (2005) and Mustajoki et al. (2007) provided decision aid to wisely choose the next swap in the MCDA method even swaps. Mustajoki et al. (2005) studied which criterion should be used as the reference criterion for the weight elicitation to harness preference programming effectively. Liesiö et al. (2007, 2008) briefly examined the problem for portfolio decision analysis but only offered limited guidance for an option elicitation order.[1]

**Table 1**
Symbols.

| Symbol | Description |
| --- | --- |
| $B$ | Budget |
| $c_i$ | Cost of option $i$ |
| $\varepsilon$ | Tolerance value |
| $i/I$ | Index/set of options |
| $i^*$ | Robust option |
| $j/J$ | Index/set of criteria |
| $k/K$ | Index/set of assessors (workshop participants) |
| $\theta$ | Parameter for social judgement scheme |
| $\varphi/\hat{\varphi}$ | Value score of the robust option in the MCDA case; borderline value-for-money ratio in the PDA case; $\hat{\varphi}$ is the heuristic choice for $\varphi$ |
| $r_i/r_i^*$ | Value-for-money ratio of option $i$; $r_i^*$ is the prediction for $r_i$ |
| $S_w$ | Set of extreme points of the convex hull of feasible weight combinations |
| $v_{ij}/v_{ij}^*/v_{ij}^a/v_{i.}^l$ | Value score of option $i$ for criterion $j$; $v_{ij}^*$, $v_{ij}^a$ and $v_{ij}^p$ are, respectively, the predicted, the actually agreed and the proposed value for $v_{ij}$ |
| $\underline{v}_{ij}/\bar{v}_{ij}$ | Lower/upper bound for $v_{ij}$ |
| $\underline{v}_{i'j}^{\#i}/\bar{v}_{i'j}^{\#i}$ | Lower/upper bound for $v_{i'j}$ after assuming that $v_{ij}$ takes its predicted value $v_{ij}^*$ |
| $v_{ijk}$ | Value for $v_{ij}$ assigned by assessor $k$ |
| $V_i/V_i^*/V_i^a$ | Overall value score of option $i$; $V_i^*$ and $V_i^a$ are, respectively, the predicted and the actually agreed value for $V_i$ |
| $\underline{V}_i/\bar{V}_i$ | Lower/upper bound for $V_i$ |
| $\underline{V}_{i'}^{\#i}/\bar{V}_{i'}^{\#i}$ | Lower/upper bound for $V_{i'}$ after assuming that $V_i$ takes its predicted value $V_i^*$ |
| $\underline{V}_i^{\#J}/\bar{V}_i^{\#J}$ | Lower/upper bound for $V_i$ after assuming that $w_j$ takes its predicted value $w_j^*$ |
| $V_{ik}$ | Value for $V_i$ assigned by assessor $k$ |
| $w_j/w_j^*/w_j^a$ | 0–1 normalised weight of criterion $j$; $w_j^*$ and $w_j^a$ are, respectively, the predicted and the actually agreed value for $w_j$. A tilde $\sim$ on top indicates that the weight is not normalised. |
| $w_{jk}$ | 0–1 normalised value for $w_j$ assigned by assessor $k$. A tilde $\sim$ on top indicates that the weight is not normalised. |
| $\underline{w}_j/\bar{w}_j$ | 0–1 normalised lower/upper bound for $w_j$. A tilde $\sim$ on top indicates that the weight is not normalised. |
| $z_{ijk}$ | Centrality position of group member $k$ for the assessment of score $v_{ij}$ according to the social judgement scheme |

The paper proceeds as follows: Section 2 briefly discusses normative preference programming techniques to identify definitely dominated options. Section 3 explains descriptive approaches developed by psychologists to predict on which precise scores $v_{ij}^*$ the group may finally agree. Section 4 pulls together the normative and descriptive decision-making perspectives from the previous two sections and develops a prescriptive measure for the Value of Agreement on option scores in MCDA. Section 5 modifies this measure for multi-criteria portfolio decision analysis (PDA)—an important extension of MCDA. Section 6 adapts the Value of Agreement to weights. Section 7 demonstrates the effectiveness of the proposed measures with a computer simulation. Section 8 reports on the application of the Value of Agreement concept to a PDA-based recruitment selection problem using computer software. Section 9 concludes with a research outlook.

**2 Preference programming**

Preference programming encompasses a set of techniques to eliminate definitely dominated options and ultimately identify the robust option given the incomplete information about scores and weights (Arbel 1989, Punkka & Salo 2013, Salo & Hämäläinen 1992, 2010). The expressed incomplete preferences are typically range-based or ordinal-based (e.g. Kirkwood & Sarin 1985, Liesiö et al. 2007, Punkka & Salo 2013). Preference programming is a normative approach that, in its original form, captures the incomplete information in a set of linear constraints. Non-dominated options can be easily identified by examining the extreme points of the resulting convex hull (e.g. Hazen 1986, Liesiö et al.

2007, White et al. 1984) or by solving mathematical programmes (e.g. Arbel & Vargas 1993, Lahdelma et al. 1998, Salo & Hämäläinen 1995).

The *range-based strict dominance rule* (e.g. Hazen 1986)

$$i_1 \succ i_2 \Leftrightarrow \underline{V}_{i_1} \geq \bar{V}_{i_2} \text{ and } \bar{V}_{i_1} > \bar{V}_{i_2}$$

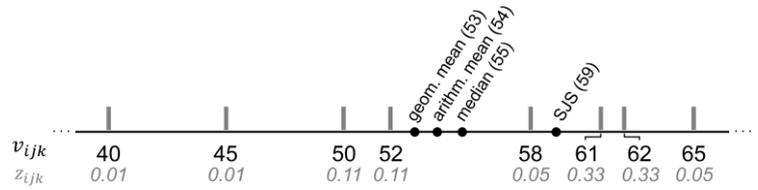

**Fig. 1.** Example of $|K| = 8$ individual appraisals $v_{ijk}$ with corresponding SJS centrality positions $z_{ijk}$ and predicted agreement $v^*_{ijk}$ according to the geometric mean, the arithmetic mean, the median and SJS.

is a logic based on preference ranges which can be applied in a basic MCDA problem where just the score information is incomplete. For the robust option $i^*$, the strict preference $i^* \succ i$ must hold for all $i \in I\setminus\{i^*\}$.[2]

An *ordinal-based weak dominance rule* can be constructed by considering participants' implicit rankings when assigning scores to options. Option $i_1$ weakly dominates option $i_2$ on criterion $j$ if all participants $k$ believe that $v_{i_1jk} \geq v_{i_2jk}$. In this case, the linear constraint $v_{i_1j} \geq v_{i_2j}$ can be added to the preference programme. Option $i_1$ weakly dominates options $i_2$ overall if all participants $k$ believe that $V_{i_1k} = \sum_j w_j v_{i_1jk} \geq \sum_j w_j v_{i_2jk} = V_{i_2k}$. In this case, the linear constraint $\sum_j w_j v_{i_1j} \geq \sum_j w_j v_{i_2j}$ can be added to the preference programme. Moreover, assume there is already a group agreement that $v_{i_1j} \geq v_{i_2j}$. In this condition, learning that the group agreed on the precise number $v^a_{i_1j}$ ($v^a_{i_2j}$) for the score $v_{i_1j}$ ($v_{i_2j}$) may allow for tightening the upper (lower) bound such that $\bar{v}_{i_2j} := \min\{v^a_{i_1j}, \bar{v}_{i_2j}\}$ ($\underline{v}_{i_1j} := \max\{\underline{v}_{i_1j}, v^a_{i_2j}\}$). The same tightening rule can be used for option scores. If there is a group consensus on $V_{i_1} \geq V_{i_2}$ and the participants agree on the precise option score $V^a_{i_1}$ ($V^a_{i_2}$), the upper (lower) bound for $V_{i_2}$ ($V_{i_1}$) can be updated to $\bar{V}_{i_2j} := \min\{V^a_{i_1j}, \bar{V}_{i_2}\}$ ($\underline{V}_{i_1} := \max\{\underline{V}_{i_1}, V^a_{i_2}\}$).

## 3 Predicting agreement

The group's actual agreement on a precise number $v^a_{ij}$ for a score $v_{ij}$ can be predicted when all group members have already made their individual appraisals $v_{kij}$. Next, we distinguish between the case in which the participants only know the single appraisals $v_{kij}$ or their ranges $[\underline{v}_{ij}, \bar{v}_{ij}]$ (section 3.1) and the case in which an aggregate of their appraisals such as the mean or the median is presented to the participants (section 3.2).

### 3.1 Social judgement scheme

The social decision scheme (Davis 1973) has led to much insight into how small groups arrive from individual preference statements to a group decision along a discrete scale of measurement (Parks & Kerr 1999). A major finding of this research stream is that individuals whose preferences are quite similar to those of other group members wield an exponentially greater influence on the final decision than those individuals with extreme preferences. Davis (1996) manifests this observation in the equations of his social judgement scheme (SJS)—a continuous generalisation of the discrete social decision scheme.

$z_{kij}$ denotes the SJS centrality position of group member $k$ for the assessment of score $v_{ij}$ and is defined as

$$z_{ijk} = \frac{\sum_{l=1, k \neq l}^{|K|} e^{-\theta|v_{ijk} - v_{ijl}|}}{\sum_{h=1}^{|K|} \sum_{l=1, h \neq l}^{|K|} e^{-\theta|v_{ijh} - v_{ijl}|}}, \theta = 1.$$

$z_{ijk}$ can be interpreted as the percentage of influence of assessor $k$ on the group's collective choice for $v_{ij}$. $v^*_{ij} = \sum z_{ijk} v_{ijk}$ is the most central value for $v_{ij}$ according to SJS, and thus a good point estimator of the group's unknown actual agreement $v^a_{ij}$. Fig. 1 exhibits an example where the point estimator $v^*_{ij}$ is derived from individual appraisals using three common aggregates and SJS.

Wide research on the social decision scheme provides a solid theoretical foundation for the centrality position approach. Furthermore, experiments by Bonner et al. (2004), Davis et al. (1997), Nadler et al. (2001) and Ohtsubo et al. (2002) offer good evidence in favour of SJS. Applications of SJS in decision support systems for MCDA workshops can be found in the studies conducted by Rigopoulos (2008), Tsiporkova & Boeva (2006) and Tundjungsari et al. (2012). It is important to also highlight that SJS shares the property of heavily discounting extreme opinions with the geometric mean as the consensus estimator. While there are many case studies that use the geometric mean in group decision problems modelled as Analytical Hierarchy Process (e.g. Saaty & Vargas 2001, Vaidya & Kumar 2006), we are not aware of any research that validates the geometric mean as an appropriate consensus estimator.

It needs to be mentioned that the strength of SJS is in majority/plurality situations where no group member can plausibly demonstrate why their judgement must be the correct one. Given that preference statements are rather subjective (e.g. Keeney 1992, Payne et al. 1993, Slovic & Lichtenstein 2006), this assumption is normally fulfilled for the MCDA problem. In situations where members with extreme opinions can (i) plausibly demonstrate the correctness of their positions, (ii) have hierarchal power or (iii) are recognised as the group's experts, the centrality concept of SJS is not the right approach (Bonner et al. 2007, see also: Laughlin 1980, Laughlin & Ellis 1986).

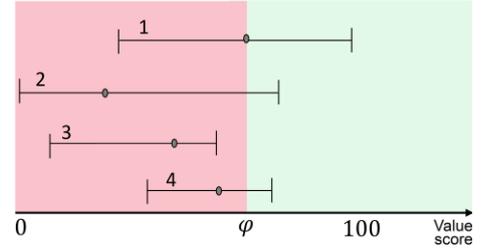

**Fig. 2.** Value score chart – bars represent feasible score interval for each option.

*3.2 Proposed agreement*

In some consensus-based MCDA workshops, the group is presented with a proposed aggregate $v_{ij}^p$ of the individual appraisals $v_{ijk}$ to support and expedite its quest for a consensus. For instance, the mean/median is often seen as a good estimator of the unknown 'optimal' value (called 'wisdom of crowds' by Surowiecki 2005, see also: Ariely et al. 2000, Sunstein 2005), from which the group should only diverge with good reasons. It is also possible to have more sophisticated proposed aggregates $v_{ij}^p$ that consider fairness concerns among group members or uneven influence of participants on the collective decision based on positional power or expertise (e.g. Keeney 2013, van den Honert 2001).

Viewing the proposed aggregate prompts an anchoring heuristic (Tversky & Kahneman 1974) and it becomes likely that the group finally agrees on this presented value. For example, we observed in a recent decision analysis workshop at the National Audit Office in Britain that the participants chose the median shown to them beforehand for 31 out of 35 options. Hence, the predicted agreement $v_{ij}^*$ should take the value of the proposed aggregate $v_{ij}^p$ in such situations.

**4 Measures for the Value of Agreement in MCDA**

Our proposed measure for the Value of Agreement rests on the assumption that one can make reasonably precise predictions $V_i^* = \sum_j w_j v_{ij}^*$ on what scores the group will agree on for options $i$, as outlined in the previous section. The score ranges $[\underline{V}_i, \overline{V}_i]$ and the predicted agreements $V_i^*$ can be graphically presented in a value score chart. It will be shown that such a chart helps define two types of measures, denoted $g_A$ and $g_B$ henceforth, indicating the likelihood of eliminating dominated options without reaching prior full agreement on them. Bringing the two measures together in a lexicographical fashion leads to a recommendation on which options the group should next seek agreement.

In a value score chart (see Fig. 2), the options are displayed in the form of horizontal lines along a scale of 0–100. The horizontal lines represent the option score ranges $[\underline{V}_i, \overline{V}_i]$ and the dots on them show the predicted agreements $V_i^*$. The borderline between the red area (appears in darker grey on black-and-white prints) and the green area (appears in light grey) marks the potentially still unknown agreed precise value score $\varphi$ for the still unknown robust option. $\hat{\varphi} = \max\{V_i^*\}$ is a heuristic choice for $\varphi$. $V_i^*$ is always replaced by $V_i^a = \sum_j w_j v_{ij}^a$ if the group has already elicited full agreement on option $i$. Note that the heuristic choice $\hat{\varphi}$ underestimates $\varphi$ systematically, as the expected value of the maximum of a set of random variables is larger than the maximum of the expected values of the same set of random variables (Chung 1948). There is a research opportunity to explore distribution functions for the agreement rather than just settling for a point estimator $V_i^*$. Such work would be interesting in itself and could also lead to unbiased heuristic choices for $\varphi$.

For those options $i$, whose bars are partially in the value score chart's green area, i.e. $\varphi \in (\underline{V}_i, \overline{V}_i)$, it is not at all certain whether one can identify them as dominated without further agreement on their value scores. To quickly eliminate dominated options, options $i$ whose hypothetical agreements $V_i^*$ would not only shift themselves, but also other options outside the green area (or on its edge[3]) due to ordinal-based weak dominance relations would therefore be particularly attractive to a group discussion. Let $[\underline{V}_{i'}^{\#i}, \overline{V}_{i'}^{\#i}]$ be the overall score range of option $i' \in I$ after having learnt that option $i$ takes the predicted value score $V_i^*$, then the count

$$g_A(i) = |I| - \left| \{i' \in I : \varphi \in (\underline{V}_{i'}^{\#i}, \overline{V}_{i'}^{\#i})\} \right|$$

is a simple measure for the Value of Agreement for option $i$. $\varphi$ takes the value of the heuristic choice $\hat{\varphi}$ in our computer simulation in section 7.3. Assume, for instance, that the participants collectively agree that option 4 is better than option 2 on a particular criterion $j$ in Fig. 2. Knowing the actually agreed criterion score $v_{4j}^a$ for option 4 may allow pushing the upper bound of option 2 leftwards when using the tightening rule from section 2, such that $\bar{V}_2^{\#4} \leq \varphi$. This, in turn would make it much more likely to rule out that option 2 requires further agreement for it to be identified as dominated. In this example, $g_A(4) = 1$.[4]

In many cases, the agreement on ordinal preference relations is not sufficient to allow shifting a further option besides $i$ off the green area after learning that $V_i = V_i^*$. A tie-breaker thus is needed: if the upper bound of option $i$ is to the far right from the borderline $\varphi$, i.e.

$$g_B(i) = \bar{V}_i - \varphi$$

is large, it is less likely that the group can avoid some further agreement on its score. Thus, it would be wise to discuss those options with large $g_B$ first. For the example in Fig. 2, $g_B$ recommends choosing option 1 next.

Pulling the two measures $g_A$ and $g_B$ together leads to the lexicographic rule

$$\underset{i \in I}{\mathrm{argmax}}\, g_A(i) \gg \underset{i \in I}{\mathrm{argmax}}\, g_B(i)$$

for selecting the next option for group discussion. In words: choose the option $i$ that maximises $g_A(i)$ and if more than one possible option $i$ exists, then select among those the option that maximises $g_B(i)$.

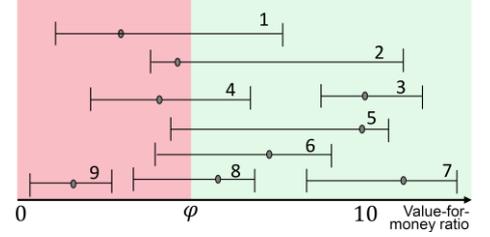

*Fig. 3.* Value-for-money chart: bars represent feasible value-for-money ratio for each option.

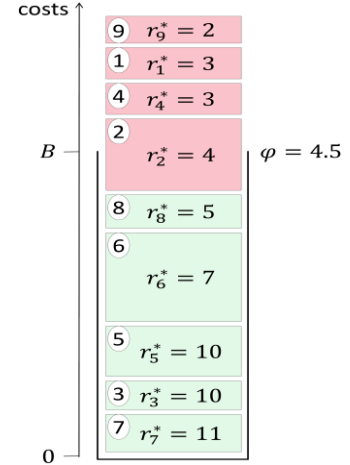

*Fig. 4.* Options ordered by decreasing predicted value-for-money ratio $r_i^*$.

## 5 Measures for the Value of Agreement in PDA

Portfolio decision analysis (PDA) extends MCDA by allowing the selection of more than one option (Phillips & Bana e Costa 2007, Salo et al. 2011). In the basic form of the PDA problem, fixed costs $c_i$ are attached to each option (typically called 'project' in PDA), options can be chosen independently of each other, and the total number of options is limited by a budget $B$. A robust portfolio $p^*$ has at least an as high an overall score of $\sum_{i \in p^*} V_i$ (and for one precise score allocation, a higher overall score) as that of any other feasible portfolio $p$ independently of which scores the group agrees on within the remaining ranges $[\underline{V}_i, \bar{V}_i]$. The PDA is a special knapsack optimisation problem (Martello & Toth 1990), in which the majority of cases favour options with a high value-for-money ratio $r_i = V_i/c_i$ (Phillips & Bana e Costa 2007). The preference programme from Liesiö et al. (2007) is currently the only available knapsack optimisation approach for incomplete score information that allows the identification of options which are definitely or definitely not in the robust portfolio. The remainder of this section will derive a measure for the Value of Agreement analogously to the MCDA case.

In a value-for-money chart (see Fig. 3), the options are displayed in the form of horizontal ratio bars representing the feasible ratios $r_i \in [\underline{V}_i/c_i, \bar{V}_i/c_i]$. When ordering the options by their value-for-money ratios, one can calculate the borderline value-for-money ratio $\varphi$, defined as the mean of the value-for-money ratio of the last option that still fits completely and the first option that fits only partially in the portfolio (see Fig. 4). A heuristic choice $\hat{\varphi}$ for the borderline ratio $\varphi$ can be computed by using the predicted value-for money-ratios $r_i^* = V_i^*/c_i$. $\varphi$ is the border between the red and green areas in the value-for-money chart.

For options $i$ crossing the borderline ratio, i.e. $\varphi \in [\underline{r}_i, \bar{r}_i]$, it is quite unclear whether they will be in the robust portfolio. Again, options whose predicted agreement $V_i^*$ would not only shift themselves, but also other options away from the borderline due to ordinal preference relations should be considered first for the group discussion. Let $[\underline{V}_{i'}^{\#i}, \bar{V}_{i'}^{\#i}]$ be the updated overall score range of option $i' \in I$ assuming that $i$ takes the predicted option score $V_i^*$, then

$$g_A(i) = |I| - \left| \left\{ i' \in I : \varphi \in \left[ \frac{\underline{V}_{i'}^{\#i}}{c_{i'}}, \frac{\bar{V}_{i'}^{\#i}}{c_{i'}} \right] \right\} \right|$$

provides a first measure for the Value of Agreement for option $i$. Assume, for example, that the collective agreement is that options 5 and 6 are both better than option 8 in Fig. 3. In such a case, knowing the agreement on option 8 makes it likely that options 5 and 6 can be ruled out as requiring further agreement to identify them as definitely in the robust portfolio. In this example, $g_1(8) = 6$.

As a tie-breaker, the distance between the borderline ratio and the nearest end of the option's ratio bar, i.e.

$$g_B(i) = \begin{cases} \min\{|\varphi - \underline{r_i}|, |\bar{r}_i - \varphi|\} &, \varphi \in [\underline{r_i}, \bar{r}_i] \\ -\min\{|\varphi - \underline{r_i}|, |\bar{r}_i - \varphi|\} &, \varphi \notin [\underline{r_i}, \bar{r}_i] \end{cases},$$

is used. If $i$'s ratio bar does not intersect the borderline ratio, $g_B(i)$ obtains a negative value. The larger the $g_B$ of an option, the less likely it is that the group can avoid some further agreement on the option's score. In Fig. 3, the $g_B$ measure recommends option 1 next.

The next option to be discussed by the group is again chosen using the lexicographic rule

$$\underset{i \in I}{\mathrm{argmax}}\, g_A(i) \gg \underset{i \in I}{\mathrm{argmax}}\, g_B(i)\,.$$

It should be noted that the two measures for the Value of Agreement described in this paper work only for the basic PDA problem. Model extensions, such as incomplete cost information, options that cannot be independently chosen or an additional constraint on the maximum number of options in the portfolio would require different measures for the Value of Agreement.

**6 Extension to weights**

In this section, we adapt the previously developed measures for the Value of Agreement to SWING weighting with incomplete weight information. The value score information $v_{ij}$ is assumed to be complete in the following calculations and assessors $k$ only appraise the weights $w_{jk}$ for all criteria $j$ individually. We also assume that the popular SWING weighting method (Edwards & Barron 1994, Mustajoki et al. 2005, von Winterfeldt & Edwards 1986) is employed and that the decision making group jointly agrees on a reference criterion as a first step. A fixed, non-normalised weight (e.g. 100 points) is assigned to reference criterion $j^r$ what establishes a linear scale of measurement. Each participant assigns non-normalised individual weights $\widetilde{w}_{jk}$ to all other criteria $j \in J\setminus\{j^r\}$, reflecting their importance relative to the reference criterion. The resulting non-normalised weight ranges span from $\underline{\widetilde{w}}_j = \min_k\{\widetilde{w}_{jk}\}$ to $\overline{\widetilde{w}}_j = \max_k\{\widetilde{w}_{jk}\}$ within which the group seeks actual agreements $\widetilde{w}_j^a$. Since the non-normalised individual weights $\widetilde{w}_{jk}$ and the actual agreements $\widetilde{w}_j^a$ are all based on the same scale of measurement, established by the reference criterion $j^r$, the Value of Agreement for SWING weights can be computed in a similar method as for scores.

The weight ranges $[\underline{\widetilde{w}}_j, \overline{\widetilde{w}}_j]$ form a convex hull of feasible weight combinations. To compute the minimum and maximum overall value score of options, only $S_w$ as the set of extreme points of this convex hull is relevant (Hazen 1986, Liesiö et al. 2007). Without loss of generality, it is assumed in the following that the weight combinations $\{w_j\}_{j \in J}$ in $S_w$ are always normalised such that they sum up to 1; i.e. $\sum_j w_j = 1\ \forall \{w_j\}_{j \in J} \in S_w$. An option's overall value score lies between

$$\underline{V}_i = \min\left\{\sum_j w_j v_{ij} : \{w_j\}_{j \in J} \in S_w\right\} \text{ and } \overline{V}_i = \max\left\{\sum_j w_j v_{ij} : \{w_j\}_{j \in J} \in S_w\right\}.$$

In the following, $w_j^*$ denotes the predicted weight agreement for criterion $j$ derived analogously to section 3 and takes the value $w_j^a$ after a group agreement has been reached. The predicted weights $w_j^*$ are normalised and sum up to 1.

Since this paper only aims to introduce the new concept Value of Agreement, applications to other weight elicitation methods are not explored further. For instance, applying the Value of Agreement to the AHP (Saaty 1990) would require prioritising the elicitation of weight ratio pairs and addressing the issue of inconsistencies. For the point assignment method (Nutt 1980), the fact that the actual agreed-upon weights must add up to a constant sum (e.g. 100 points) and therefore cannot be elicited independently from each other poses a challenge to the feasibility of applying the Value of Agreement concept. Finally, if the decision makers choose to structure a larger number of criteria in a value tree (Keeney & Raiffa 1976), any Value of Agreement measure would have to take into account the interdependencies of the weights on different hierarchy levels.

## 6.1 Preference programming

As a range-based strict dominance rule, we use the linear programming approach from Hannan (1981) for the MCDA case and the dynamic programming approach from Liesiö et al. (2007)[5] for the PDA case. The ordinal-based weak dominance rule for weights $w_j$ is analogous to value scores $v_{ij}$ (see section 2).

## 6.2 MCDA

The heuristic choice for the borderline $\varphi$ is $\max_i\{\sum_j w_j^* v_{ij}\}$. Let $[\underline{V}_i^{\#j}, \bar{V}_i^{\#j}]$ be the overall score range of option $i \in I$ after having learnt that criterion $j$ takes the predicted weight $w_j^*$ and having applied the ordinal-based weak dominance rule. The first measure for the Value of Agreement,

$$g_A(j) = |I| - \left|\{i \in I : \varphi \in (\underline{V}_i^{\#j}, \bar{V}_i^{\#j})\}\right|,$$

counts the number of options whose upper bounds are still larger than the heuristic choice after assuming $w_j = w_j^*$. In general, $g_A$ favours 'high impact' criteria $j$ with a large normalised weight span $\bar{w}_j - \underline{w}_j$ and large variance $\text{Var}_i\{v_{ij}\}$ among its criterion scores. Using

$$g_B(j) = \max_i\{\bar{V}_i^{\#j} - \varphi\}$$

as a tie-breaker, the lexicographic rule for selecting the next criterion to be discussed by the group is

$$\underset{j \in J}{\text{argmax}}\, g_A(j) \gg \underset{j \in J}{\text{argmax}}\, g_B(j).$$

## 6.3 PDA

The borderline value-for-money ratio $\varphi$ is computed as described in section 5 using $V_i^* = \sum_j w_j^* v_{ij}$ as the predicted agreement. $[\underline{V}_i^{\#j}, \bar{V}_i^{\#j}]$ is the updated overall score range after assuming $w_j = w_j^*$. The first measure for the Value of Agreement is

$$g_A(j) = |I| - \left|\left\{i \in I : \varphi \in \left[\frac{\underline{V}_i^{\#j}}{c_i}, \frac{\bar{V}_i^{\#j}}{c_i}\right]\right\}\right|,$$

again favouring 'high impact' criteria. The normalised weight range is used as a tie-breaker; i.e.

$$g_B(j) = \bar{w}_j - \underline{w}_j.$$

Again, the lexicographic rule

$$\underset{j \in J}{\text{argmax}}\, g_A(j) \gg \underset{j \in J}{\text{argmax}}\, g_B(j)$$

is used to choose the next criterion to be discussed by the group.

## 7 Computer simulation

### 7.1 Test dataset

Several authors (e.g. Keisler 2004, Mustajoki et al. 2005, Salo & Hämäläinen 2001, Vetschera et al. 2014) have previously deployed extensive computer simulation as a means to test ideas on preference programming and value of information. To demonstrate the effectiveness of the proposed measures for the Value of Agreement, a large test dataset with 360 instances was generated (see Table 2 and Supplementary Material A). The dataset comprises 20 instances for each possible combination of number of options, score distribution, weight distribution and cost distribution as provided in Table 2.

The group members' individual appraisals $v_{ijk}$ are constructed as follows:

- Two participants $k_1 \in K$ and $k_2 \in K$ are always randomly chosen and are assigned the upper and lower criterion score bound; i.e. $v_{ijk_1} = \underline{v}_{ij}$ and $v_{ijk_2} = \bar{v}_{ij}$.
- $v_{ijk}$ is normally distributed with $\mathcal{N}\left(\frac{\underline{v}_{ij}+\bar{v}_{ij}}{2}, \left(\frac{\underline{v}_{ij}+\bar{v}_{ij}}{8}\right)^2\right)$ for $k \in K\backslash\{k_1, k_2\}$ and rounded to the nearest integer. If $v_{ijk} \notin [\underline{v}_{ij}, \bar{v}_{ij}]$, new numbers are generated for it until $v_{ijk} \in [\underline{v}_{ij}, \bar{v}_{ij}]$.

The group members' actual agreements $v_{ij}^a$ are modelled by two different distributions A1 and A2. The more predictable median-focused distribution A1 is triangular with $T(\underline{v}_{ij}, v_{ij}^*, \bar{v}_{ij})$. If the originally predicted agreement $v_{ij}^*$ is smaller than $\underline{v}_{ij}$ or larger than $\bar{v}_{ij}$ due to bound tightening, it takes the value of the bound $\underline{v}_{ij}$ or $\bar{v}_{ij}$, respectively. The less predictable distribution A2 is uniform with $U(\underline{v}_{ij}, \bar{v}_{ij})$. For each combination of properties from Table 2, 10 instances receive actual agreement scores according to A1 and the other 10 according to A2. Starting with $v_{11}$, the actual agreement scores $v_{ij}^a$ are computed one after the other. After every assignment of a new actual agreement score $v_{ij}^a$, the feasible ranges $[\underline{v}_{i'j}, \bar{v}_{i'j}]$ for all still non-précised scores are tightened as described in section 2. After full actual agreement $V_i^a$ about a further option $i$ is constructed, the ranges $[\underline{V}_{i'}, \bar{V}_{i'}]$ of all still non-précised options are tightened as described in section 2. If $V_i^a$ is outside the tightened score ranges for option $i$, the whole procedure is restarted from $v_{11}^a$. The actual weight agreement $w_j^a$ is modelled analogously.

*Table 2*
Properties of test dataset.

| | | |
|---|---|---|
| $\|K\|$ | **8** | |
| $\|I\|$ | **25, 50** and **100** | |
| $w_j$ | **W0:** 5 criteria with random weights $\ddot{w}_j$ so that $\sum_j \ddot{w}_j = 1$ and $\underline{w}_j = \ddot{w}_j = \bar{w}_j \,\forall j$ (complete weight information) <br> **W1:** 5 criteria with random weights $\ddot{w}_j$ so that $\sum_j \ddot{w}_j = 1$. $\underline{w}_j$ and $\bar{w}_j$ take with equal probability one of the following five value pairs: $\underline{w}_j + 0.02 = \ddot{w}_j = \bar{w}_j - 0.03$; $\underline{w}_j + 0.05 = \ddot{w}_j = \bar{w}_j - 0.05$; $\underline{w}_j + 0.07 = \ddot{w}_j = \bar{w}_j - 0.08$; $\underline{w}_j + 0.1 = \ddot{w}_j = \bar{w}_j - 0.1$; and $\underline{w}_j + 0.12 = \ddot{w}_j = \bar{w}_j - 0.13$. | |
| $v_{ij}$ | **V0:** $\underline{v}_{ij} \sim U(0,100)$ and $\underline{v}_{ij} = \bar{v}_{ij}$     $\forall i, j$ (complete score information) <br> **V1:** $\underline{v}_{ij} \sim U(0,95)$   and $\underline{v}_{ij} + 5 = \bar{v}_{ij}$   in 25% of the cases <br>        $\underline{v}_{ij} \sim U(0,90)$   and $\underline{v}_{ij} + 10 = \bar{v}_{ij}$   in 25% of the cases <br>        $\underline{v}_{ij} \sim U(0,85)$   and $\underline{v}_{ij} + 15 = \bar{v}_{ij}$   in 25% of the cases <br>        $\underline{v}_{ij} \sim U(0,80)$   and $\underline{v}_{ij} + 20 = \bar{v}_{ij}$   in 25% of the cases <br> **V2:** $\underline{v}_{ij} \sim U(0,90)$   and $\underline{v}_{ij} + 10 = \bar{v}_{ij}$   in 25% of the cases <br>        $\underline{v}_{ij} \sim U(0,80)$   and $\underline{v}_{ij} + 20 = \bar{v}_{ij}$   in 25% of the cases <br>        $\underline{v}_{ij} \sim U(0,70)$   and $\underline{v}_{ij} + 30 = \bar{v}_{ij}$   in 25% of the cases <br>        $\underline{v}_{ij} \sim U(0,60)$   and $\underline{v}_{ij} + 40 = \bar{v}_{ij}$   in 25% of the cases | |
| $c_i$ | Let $r_i = \frac{1}{c_i'} \sum_j w_j \frac{\underline{v}_{ij} + \bar{v}_{ij}}{2}$ be the average value-for-money ratio of option $i$ and $U(\,)$ a uniform distribution, then <br> **C1:** $r_i \sim U(1, 2.5) \,\forall i$ <br> **C2:** $r_i \sim U(1, 10) \,\forall i$. <br> The resulting $c_i'$ from settings C1 and C2 are normalised to $c_i$ so that $\sum_i c_i = 2{,}500$. $B = 1{,}000$ is assumed as the budget for all problem instances. | |

*7.2 Comparison measures*

To evaluate the effectiveness of the proposed measure for the Value of Agreement, we compare it with the simple rules of MaxRange and Random. The MaxRange rule (cf. Hämäläinen & Pöyhönen 1996, Salo & Hämäläinen 1995) enquires the precise actual agreement of the option (criterion) with the greatest current score range $[\underline{V}_i, \bar{V}_i]$ (weight range $[\underline{w}_j, \bar{w}_j]$) until the robust option or portfolio is elicited. The Random rule retrieves the precise actual agreement for a randomly chosen option (criterion) for which $V_i^a$ ($w_j^a$) has not yet been revealed. Preference programming[6] is applied for both rules after every single actual agreement to exclude options that do not require further preference elicitation.

By enumerating all feasible elicitation orders, it is possible to compute a lower bound for the number of required elicitations for each test instance. This lower bound could always be reached if one knew the participants' actual agreements in advance. As the number of criteria is small in the test dataset ($|J| = 5$), a full enumeration can be performed for all instances with incomplete weight information. Designing a branch-and-bound enumeration algorithm to compute the lower bound efficiently for all instances with incomplete score information might be possible, but beyond the scope of this paper.

*7.3 Results*

The test algorithm always selects the option or criterion with the greatest Value of Agreement, reveals its modelled actual group agreement, and applies preference programming[6] on the new situation. The median is used as predicted agreement for scores and weights. Tables 3 to 6 show the results for the test dataset.

In the MCDA case, the Value of Agreement rule on average reduces the number of option (criterion) elicitations statistically significantly by 50% (8%) and 45% (19%) compared to the MaxRange and the Random rules, respectively. For incomplete score information, MaxRange consistently ignores options with small ranges on the right-hand side of the value score chart despite the fact that these are more likely to be the robust option. Therefore, the Random rule outperforms the MaxRange rule. In the PDA case, the Value of Agreement rule on average decreases the number of inspected options (criteria) statistically significantly by 41% (2%) and 47% (12%) relative to the MaxRange and the Random rules, respectively.

*Table 3*
Average number of required option agreements $V_i^a$ in the MCDA case with Value of Agreement rule (first number), MaxRange rule (second number) and Random rule (third number) for the 240 W0 test instances grouped by number of options, value score distribution and agreement model.

|     | V1 | | V2 | |
| --- | --- | --- | --- | --- |
|     | A1 | A2 | A1 | A2 |
| 25  | 1.3 \| 2.3 \| 2.1 | 1.9 \| 3.1 \| 2.7 | 2.4 \| 4.6 \| 4.7 | 2.9 \| 5.3 \| 4.8 |
| 50  | 1.6 \| 3.2 \| 2.7 | 1.6 \| 3.1 \| 2.4 | 2.9 \| 6.3 \| 5.8 | 3.1 \| 7.3 \| 6.8 |
| 100 | 2.0 \| 4.3 \| 3.9 | 2.1 \| 3.5 \| 3.2 | 2.1 \| 7.3 \| 6.4 | 3.3 \| 7.7 \| 6.3 |

*Table 4*
Average number of required option agreements $V_i^a$ in the PDA case with Value of Agreement rule (first number), MaxRange rule (second number) and Random rule (third number) for the 240 W0 test instances grouped by number of options, value score distribution, cost distribution and agreement model.

|     |     | V1 | | V2 | |
| --- | --- | --- | --- | --- | --- |
|     |     | A1 | A2 | A1 | A2 |
| 25  | C1  | 8.3 \| 12.2 \| 12.1 | 8.3 \| 10.9 \| 11.5 | 9.3 \| 14.2 \| 15.5 | 11.1 \| 13.4 \| 15.4 |
| 25  | C2  | 4.0 \| 4.8 \| 5.0 | 2.2 \| 2.8 \| 3.3 | 6.5 \| 9.3 \| 9.8 | 4.9 \| 7.2 \| 7.5 |
| 50  | C1  | 12.7 \| 24.8 \| 28.0 | 16.4 \| 27.1 \| 29.4 | 15.0 \| 27.1 \| 29.9 | 20.5 \| 31.9 \| 34.8 |
| 50  | C2  | 5.0 \| 7.4 \| 8.5 | 8.6 \| 12.0 \| 13.3 | 7.9 \| 17.4 \| 17.2 | 10.3 \| 16.5 \| 19.3 |
| 100 | C1  | 17.8 \| 49.9 \| 57.2 | 26.7 \| 55.5 \| 64.0 | 24.0 \| 50.3 \| 55.5 | 31.7 \| 53.3 \| 61.9 |
| 100 | C2  | 9.8 \| 26.7 \| 32.3 | 11.1 \| 24.2 \| 26.7 | 10.7 \| 31.5 \| 35.6 | 21.3 \| 37.8 \| 46.7 |

*Table 5*
Average number of required criterion agreements $w_j^a$ in the MCDA case for the lower bound (first number), with the Value of Agreement rule (second number), with the MaxRange rule (third number) and with the Random rule (fourth number) for the 120 V0-W1 test instances grouped by number of options and agreement model.

|     | A1 | A2 |
| --- | --- | --- |
| 25  | 1.9 \| 2.2 \| 2.2 \| 2.9 | 1.9 \| 2.2 \| 2.5 \| 2.9 |
| 50  | 1.3 \| 1.8 \| 1.8 \| 1.9 | 1.6 \| 1.9 \| 2.3 \| 2.3 |
| 100 | 1.5 \| 1.7 \| 1.8 \| 2.5 | 2.0 \| 2.4 \| 2.6 \| 2.4 |

*Table 6*
Average number of required criterion agreements $w_j^a$ in the PDA case for the lower bound (first number), with the Value of Agreement rule (second number), with the MaxRange rule (third number) and with the Random rule (fourth number) for the 120 V0-W1 test instances grouped by number of options, cost distribution and agreement model.

|     |     | A1 | A2 |
| --- | --- | --- | --- |
| 25  | C1  | 3.3 \| 3.4 \| 3.4 \| 4.3 | 3.6 \| 3.8 \| 3.9 \| 4.0 |
| 25  | C2  | 2.3 \| 2.7 \| 2.6 \| 3.3 | 2.6 \| 3.1 \| 3.1 \| 3.2 |
| 50  | C1  | 3.9 \| 4.2 \| 4.5 \| 4.6 | 4.0 \| 4.1 \| 4.2 \| 4.9 |
| 50  | C2  | 2.6 \| 3.1 \| 3.2 \| 4.0 | 3.1 \| 3.5 \| 3.6 \| 4.0 |
| 100 | C1  | 4.4 \| 4.5 \| 4.5 \| 5.0 | 4.6 \| 4.6 \| 4.7 \| 4.8 |
| 100 | C2  | 4.3 \| 4.3 \| 4.4 \| 4.8 | 3.9 \| 3.9 \| 4.1 \| 4.7 |

The results underscore that a more accurate prediction of the agreed value score improves the effectiveness of the Value of Agreement rule. For A2 instances, the Value of Agreement rule decreases the number of required option agreements by 41% in the MCDA and by 43% in the PDA case, compared to the Random rule; but for A1 instances, these numbers rise to 49% in the MCDA and 51% in the PDA case. Overall, the Value of Agreement measure seems to make the greatest gains by looking for options that trigger ordinal-based weak dominance rules, have a large score range and are not too far away from the true borderline $\varphi$. The gains from a more precise prediction of the agreement (i.e. changing from A1 to A2 instances) were much smaller. This is because prediction errors are limited by the scores' upper and lower bounds, what means that the predicted borderline $\hat{\varphi}$ should even for less accurate predictions of the agreed scores not be located too far from the unknown true borderline $\varphi$ in most cases.

Increasing the score ranges from setting V1 to V2 leads to 66% and 33% more required option agreements for the MCDA and PDA case, respectively, when using the Value of Agreement rule. This is an expected result because the more the participants disagree on the scoring, the longer it should take to reach group agreement on the robust option. Decreasing the spread of the options' value-for-money ratios from setting C2 to C1 doubles the average number of required option agreements in the PDA case when using the Value of Agreement rule. This result is also less surprising because if all options tend to be, on average, more or less equally attractive in terms of their value-for-money ratios, it becomes less likely that many options $i$ are dominated or dominating, independent of what actual scores $v_{ij}^a$ the group finally agrees on.

Unlike for incomplete score information, the reduction in required preference elicitations is small for incomplete weight information—both in percentage and nominal terms. This can be explained by the fact that the MaxRange rule already provides solutions near the lower bound. The required number of elicitations when using the MaxRange rule is just 29% and 8% higher for the MCDA and the PDA case, re-spectively, as compared to the lower bound. The Value of Agreement rule reduces this deviation from the lower bound to 19% and 6%, respectively.

**8 Application to recruitment**

Ennovators Ltd. is a medium size e-commerce brand developer with approximately 100 employees in its London office. In summer 2015, the company's marketing department advertised to recruit a pricing analyst. The recruitment exercise was conducted by a panel consisting of the head of e-commerce, a marketing manager and a pricing analyst. A decision analysis model informed the recruitment process.

Several real-world applications of decision analysis to the recruitment process have been previously reported in the literature. Most publications showcase a fuzzy approach (e.g. Afshari et al. 2013, Petrovic-Lazarevic 2001, Polychroniou & Giannikos 2009); often combined with other techniques such as AHP (Celik et al. 2009, Khosla et al. 2009), ANP (Kabak et al. 2012) and TOPSIS (Kelemenis & Askounis 2010, Wang et al. 2006). Using fuzzy sets, the authors seek to consider the uncertainties associated with appraising candidates. Stal-Le Cardinal et al. (2011) applied Electre Tri to select a portfolio of students for university admission. Gibney & Shang (2007) used AHP for recruiting the dean of a business school. Finally, Gardiner & Armstrong-Wright (2000) provided detailed insights into developing a simple MDCA model for faculty recruitment in view of anti-discrimination laws. In line with Barclay (2001), we could not find evidence in the literature showing that decision analysis is used much in practice to inform the recruitment process. Among other reasons (Jessop 2004), we argue that the complexity of the elicitation process in many previous case studies severely hinders the uptake by practitioners (cf. Belton & Stewart 2002, Keeney & von Winterfeldt 2007). Therefore, it was our aim to design a user-friendly, time-efficient and rigorous decision analysis approach for the Ennovators' recruitment process.

As the initial step of the Ennovators' recruitment process, the panel jointly constructed the decision model by agreeing on criteria, value scales and weights. Based on the job description and the company's core behavioural values, 'analytics software', 'numerical skills', 'personality fit', 'pricing experience' and 'self-motivation' were chosen as the assessment criteria. For each criterion, a candidate's minimum required performance level in order to be considered for the job was assigned a value score of 0 points and 100 points for the maximum desired performance level. Verbal descriptions of the performance levels were assigned to the value score points of 25, 50 and 75 for each criterion using the bi-section method (Fishburn 1967). The resulting value scales are shown in Table 8. Note that the verbal descriptions of the five equal-distanced points on the value scales only serve as a reference, and panel members were free to assess candidates with any other value score between 0 and 100 for a particular criterion. All five criteria are intangible in the sense that a candidate's precise value score for a criterion depends on the assessors' subjective interpretations of CV, cover letter and interview performance based on the assessors' knowledge and experience. Due to the diversity of job candidates, it is not practical to construct more precise value scales (Gardiner & Armstrong-Wright 2000). Criteria trade-offs were assessed using the SWING weighting method (Edwards & Barron 1994, von Winterfeldt & Edwards 1986) which compares the value of improvement from 0 to 100 for each criterion. The recruitment process takes the form of a portfolio decision analysis problem with a predetermined number of candidates being invited to the next interview round (the budget) and each interview causing fixed costs of 1.

An easy-to-use software was developed to support preference programming under incomplete score information for portfolio decision analysis (freely available from Supplementary Material B). Fig. 5 shows a screenshot of the software's window for the value-for-money chart from the case study. Candidates who will not and will definitely be invited for an interview are displayed on the bottom left and right, respectively. The Value of Agreement concept suggests discussing Marco next, whose value score bar[7] is highlighted. Double-clicking on the name opens a new window in which the decision makers can see the score ranges $[\underline{v}_{ij}\,\bar{v}_{ij}]$ and they can type in their jointly-agreed scores $v_{ij}^a$ for Marco. The software tool allows the choice between the social judgement scheme (used in this case study) and the

*Table 7*
Assessment criteria with weights in parenthesis and value scale.

**Analytics software (21%)**
- 100 Advanced skills in Excel (e.g. pivot tables, conditional formatting) AND considerable experience in VBA AND experience in Google Analytics AND considerable experience in at least one other relevant analytics tool (e.g. data mining software, R, SQL, computer programming)
- 75 Advanced skills in Excel AND experience in VBA AND experience in at least one other relevant analytics tools
- 50 Advanced skills in Excel AND some experience in VBA
- 25 Advanced skills in Excel
- 0 Basic skills in Excel (e.g. chart building, cell functions)

**Numerical skills (24%)**
- 100 Degree in a highly numerical subject (e.g. math, science, finance, economics) AND at least one year's work experience in an analyst role
- 75 Degree in a highly numerical subject OR at least one year's work experience in an analyst role
- 50 Degree in business administration, marketing or a related field OR some experience in applying numerical skills in a work environment
- 25 Degree which includes some numerical modules (e.g. psychology)
- 0 Non-numerical university degree

**Personality fit[8] (15%):** Creative, good command over English, experience working at a SME, open-minded, outgoing, pro-active, sense of humour
- 100 Candidate is expected to fit perfectly into the culture of Ennovators AND most of the staff are very much expected to like to work with the candidate (evidence to look for: previous work experience of dealing with people, examples of behaviour provided in cover letter and interview, and verbal and non-verbal communication in the interview)
- 75 Candidate is expected to fit well into the culture of Ennovators with some minor limitations (e.g. minor language barriers, no evidence of dealing successful with conflicts, a bit shy)
- 50 Candidate is not expected to contribute positively or negatively to Ennovators' working environment AND candidate is unlikely to cause any communication or behavioural problems
- 25 Other employees are expected to be able to work together with the candidate without any major issues AND candidate has room for improving communication and social interactions (e.g. very shy, less efficient communication)
- 0 Staff at Ennovators could work with the candidate (e.g. rather poor English, rather unskilled communicator, slightly arrogant)

**Pricing experience (29%)**
- 100 Experienced in developing pricing models or strategies
- 75 Familiar with pricing through work experience
- 50 Good understanding of pricing theory from university
- 25 Some basic understanding of pricing (e.g. through marketing modules at university)
- 0 No previous knowledge

**Self-motivated and willingness to learn (12%):** Career progression, extracurricular activities, good grades, interest in job
- 100 Top grades at university (e.g. distinction) AND strong career progression (if applicable) AND further training (e.g. master's, diplomas, languages, internships, volunteering) AND strong examples of work ethos in the cover letter or interview AND shows a satisfactory level of interest in job vacancy
- 75 Good grades (e.g. merit) AND (strong career progression OR further training) AND very good examples of work ethos AND shows a satisfactory level of interest in job vacancy
- 50 (Acceptable career progression OR some further training) AND good examples of work ethos AND shows a satisfactory level of interest in job vacancy
- 25 Poor grades (e.g. pass) AND good examples of work ethos AND shows a satisfactory level of interest in job vacancy
- 0 Shows a satisfactory level of interest in job vacancy

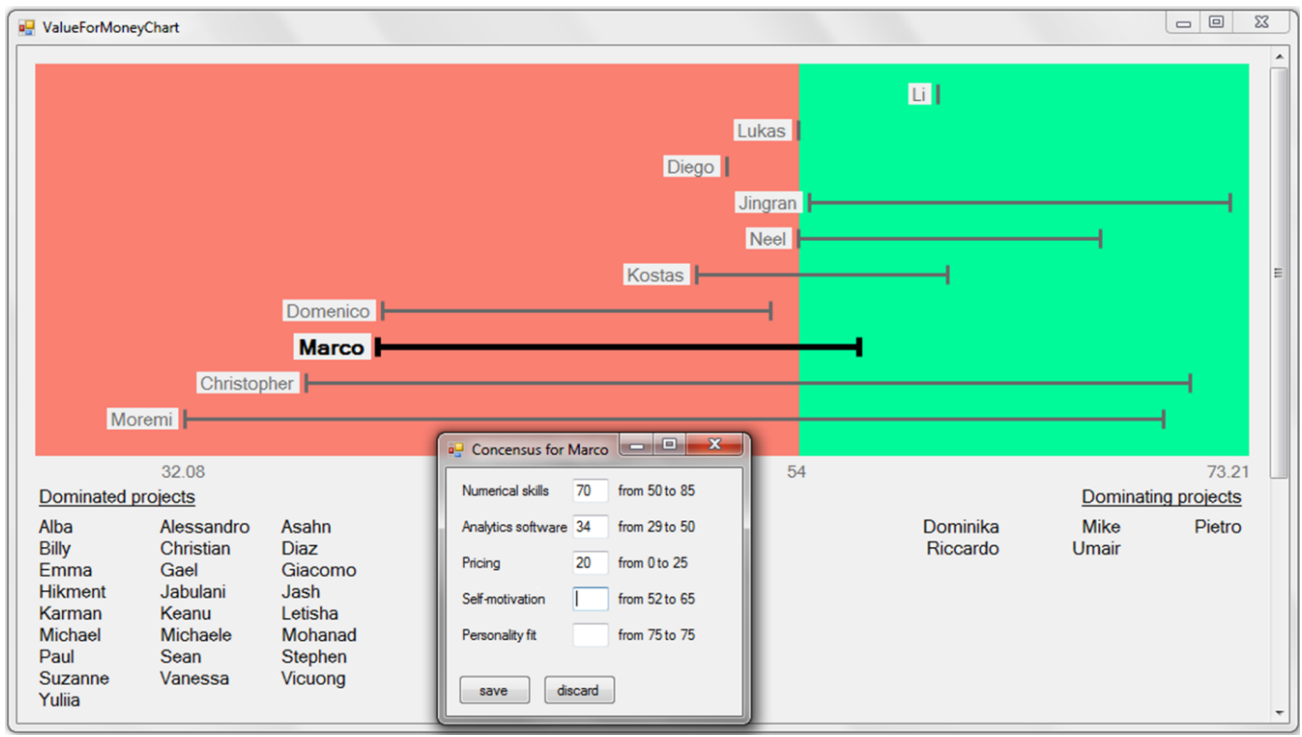

**Fig. 5.** Screenshot of value-for-money chart from the first group discussion. Candidate names have been changed.

median to predict the agreed value score. To avoid biasing the decision makers, the software's value-for-money chart does not reveal the predicted scores $V_i^*$ to the participants.

Ennovators' candidate review consisted of four steps. In step one, HR screened the 87 incoming applications and excluded 47 (e.g. no proper cover letter, requiring work permit, no university degree, obvious misfits). In step two, each panel member scored the remaining 40 applications individually using the agreed value scales in Table 7. In a subsequent one-hour group discussion, the panel members sought the necessary agreement on the scores to select the 10 candidates who should be invited to an interview (individual and agreed value scores are available in Supplementary Material C). The discussion was facilitated by a decision analyst who also entered the agreed scores into the software tool. At no point during the discussion did the panel members have a particular preference on which candidate's score agreement should be sought next, and therefore they always followed the suggestion made by the software tool. In step three, nine interviews were held soon afterwards (one candidate sent their apologies) and each panel member individually updated their previous scorings. In the subsequent group discussion, the panel members used the software tool to agree on the three candidates to be invited to the final round on the following day; this discussion was however conducted without the support from a decision analyst. In step four, a problem-solving interview was held and the panel members updated their previous individual scorings of the candidates before discussing them as a group. The software tool was not employed in step four because the panel required a full ranking of the three candidates in case the best one rejected the job offer. Overall, the panel members were confident that they made the job offer to the best candidate, which meant that the decision model was requisite (Phillips 1984).

The case study revealed two technical problems that were not relevant in the computer simulation. The first problem involves candidates on whom the panel members disagree with regards to their individual scorings; i.e. whether or not a particular candidate meets the minimum required performance level for a particular criterion. If a panel member $k$ believes that a candidate $i$ does not meet the minimum expected performance levels in criterion $j$, they would choose a prohibitive high value score of $v_{ijk} = -300$ such that candidate $i$ can practically never become a 'dominating option' unless all the panel members jointly agree on a better score for $v_{ij}^a$ following a group discussion. The second problem entails situations where the group agrees on a value score $v_{ij}^a$ for option $i$ in criterion $j$ which is outside its range $[\underline{v}_{ij}, \bar{v}_{ij}]$. This situation occurred once in the case study when none of the panel members realised before the group discussion that one candidate's personality fit was actually very poor. Some previously determined dominance relations may become invalid if the agreed value score of $v_{ij}^a$ lies outside the range $[\underline{v}_{ij} \bar{v}_{ij}]$—thus, options previously identified as dominated or dominating need to be reconsidered. In cases where an option $i$'s agreed value score on a particular criterion $j$ lies outside its original value score range, $v_{ijk}$ takes the value $v_{ij}^a$ for all participants $k$ and the preference programme is re-initialised, beginning with the complete original set of options.

Reflecting on the recruitment process, the panel members highlighted that the use of decision analysis forced them to properly assess candidates against all job requirements and to try to avoid becoming biased by individual aspects (see also Gardiner & Armstrong-Wright 2000, Gibney & Shang 2007). Although considerably more time-consuming, the panel members agreed that using proper multi-criteria decision models should lead to better and more transparent recruitment decisions. Since scoring models (in various forms and shapes) are commonly used in a business environment, the panel members readily grasped and accepted the core elements of the developed decision model (additive value function, bisection method, and SWING weighting).

The panel members also believed that the often-employed aggregation method to reach consensus in multi-criteria group decision making would have been inappropriate in this case study. As mentioned in the introduction section of this paper, the aggregation method computes a proposed aggregate $v_{ij}^p$ based on group members' individual scores $v_{ijk}$ and automatically decides on the group's preferred portfolio using the scores $v_{ij}^p$. But several times, only one panel member $k'$ made a particular observation about a candidate (e.g. noticing an important but hidden detail in the CV or knowing a particular analytics software that the candidate mentioned) which led to a rather extreme criterion score $v_{ijk'}$; however, this turned out to be justified in the subsequent group discussion in the sense that the actually agreed score $v_{ij}^a$ was close to $v_{ijk'}$ and far from the predicted score $v_{ij}^*$. Thus, calculating the preferred candidate portfolio based on automatically calculated aggregates could have led to a poor recruitment decision in this case study. On the other hand, it would not have been practically feasible to discuss all 40 applications one-by-one due to time constraints. Therefore, preference programming offered a way to retain valuable individual scores by not settling for automatic aggregates but also keeping the decision problem at a practically manageable size. In this context, the Value of Agreement can be seen as a useful tool to make preference programming more effective but also more user-friendly in situations where a skilled facilitator is not present.

## 9 Conclusion

Deciding on which option a group should discuss next in a decision analysis workshop following individual option appraisals has always involved only informed guess-work. This paper introduced the Value of Agreement concept to measure for which options it is particularly worth seeking further group agreement in order to make a robust group decision quickly. An extensive computer simulation and a case study demonstrated the usefulness of the proposed Value of Agreement concept. In addition, we provided a free software programme to allow practitioners and researchers to readily employ preference programming and the Value of Agreement for standard portfolio decision analysis.

The Value of Agreement concept is primarily developed for decision problems where the number of options is large, but the available time to seek group agreement is rather limited. Practical examples include recurring decision problems such as hiring processes, maintenance prioritisations, project tenders, credit approvals and research funding decisions. As most decision problems are assessed with only a few criteria, we are less convinced of the practical relevance of applying the Value of Agreement to weights. Furthermore, by comparing the results for three different prioritisation rules (Value of Agreement, MaxRange and Random) with the theoretical lower bound, the possibilities for reducing the number of required elicitations were shown to be limited for incomplete weight information.

We broadly defined the Value of Agreement as an estimate for the impact of eliciting additional preference information on the time requirements of a decision analysis workshop. The present paper applies the new concept only to a specific workshop design and problem formulation. Therefore, there is scope for more research. First, the present paper assumes that the group always seeks maximal agreement on an option's value score $V_i$ before discussing the next option. Future work may advocate a more flexible approach of switching between discussing the individual criterion value scores $v_{ij}$ from different options. In this case, the Value of Agreement concept could be used to provide guidance on which criterion value scores $v_{ij}$ seeking further agreement is particularly worthwhile and when it might be advisable to move on to the next option without having reached full agreement on the option's value score $V_i$. Thereby facilitators need to be mindful of the additional cognitive burden to participants that might be imposed by the frequent switching of the discussed option. Second, the present paper assumes that a group is prepared to discard an option only if it is strictly dominated. However, in some cases, the maximum loss of value in discarding a particular option may be so small that it would not be worth the decision making group's time (Salo & Hämäläinen 2010). Thus, future applications of the Value of Agreement concept might explicitly consider the costs of decision makers' time when proposing the next option for group discussion. Ultimately, the Value of Agreement concept might be used to seek group consensus on preference assessments until the maximum loss of value of the decision problem has been reduced to a level that the group feels comfortable with. The final decision could be made later by using the aggregation method mentioned in the introduction section, by applying decision rules for incomplete preferences such as minimax regret or by delegating the final decision to particular individuals for further considerations. Third, the present paper assumes that decision makers express their preferences numerically. Alternatively, the Value of Agreement concept can also be applied to qualitative preference judgements (e.g. Fasolo & Bana e Costa 2014). Those qualitative judgements can be in the form of preference classes (e.g. 'very good'), ordinal preference relations (e.g. 'X belongs to the 5 best options') or intensity preferences (e.g. 'weakly preferred'). A behind-the-scenes model translates the qualitative judgements into incomplete preference information (e.g. Bana e Costa & Vansnick 1994, Punkka & Salo 2013) and solves the resulting preference programme. Again, the question is where the decision maker should begin to make their qualitative judgements more precise. Fourth, the present paper applies the Value of Agreement concept only to SWING weighting. If future research can demonstrate that the Value of Agreement is indeed practically useful for weights, the adaption of this concept to other weight-elicitation methods would be an obvious research opportunity. Fifth, the present paper only addresses a basic PDA formulation. Applications of the Value of Agreement concept to more general PDA cases would be welcomed.

Future work may also develop measures for the Value of Agreement in a more systematic way than in this paper. Better heuristic choices for the predicted score of the robust option in the MCDA case and better usage of the knapsack problem characteristic in the PDA case would be desirable. It is also possible to examine whether there are benefits from recommending the next option for group discussion based on more complex option agreement scenarios. For instance, a decision tree could be used to model how the choice for the next option is influenced by which options the group might select afterwards and what happens if the group agreement deviates from the prediction. Finally, a more axiomatic approach to the constructions of measures for the Value of Agreement, with proven properties of the solution quality, would be of theoretical interest.

As a final point, it should not be concluded from this paper that the group must always mechanically attempt to find full agreement on the option with the greatest Value of Agreement. Sometimes there are options for which agreement can be found very easily or those on which the participants have fundamentally different perspectives (Mustajoki & Hämäläinen 2005). Dias (2007) argues that the participants should first focus on the less difficult elicitation questions to allow them to learn more about the used method and their preferences before tackling the difficult ones. One may also want to consider that gaining almost no progress in finding agreement from the start can endanger the 'emotional life' (Phillips & Phillips 1993) of the group and thus, the success of the workshop. Therefore a skilled facilitator should balance the recommendations prescribed by the Value of Agreement and the need to tackle easy tasks first.

**Acknowledgement:** The author is grateful to Dr Gilberto Montibeller for suggesting the research topic and Trang Le for making the case study at Ennovators Ltd. possible. He also thanks Prof Ahti Salo and Dr Andrew Wilshere for their helpful feedback.

**Endnotes**

[1] Following their argument, options with a core index of around 0.5 should probably be examined first.

[2] As an alternative to Hazen's strict dominance rule, one may apply a quasi-dominance rule instead (Dias & Climaco 2005). In this case, an option $i^*$ is defined as robust if no other option $i \in I \setminus \{i^*\}$ with an upper bound $\bar{V}_i$ exceeding the lower bound $\underline{V}_{i^*}$ of option $i^*$ by more than a given tolerance value $\varepsilon$ exists. Quasi-dominance allows taking into account that the maximum loss of value the group could suffer when choosing option $i^*$ may be too small to justify letting the decision making group convene for longer (Salo & Hämäläinen 2010). Quasi-dominance would add more complexity to the decision model and therefore is not discussed further in this paper.

[3] When the heuristic choice $\hat{\varphi}$ takes the value of an already agreed overall value score of an option $s$ (i.e. $\hat{\varphi} = V_s^a$), the upper bounds of some other options $t$ might be tightened such that $\bar{V}_t = V_s^a = \hat{\varphi}$. Those options $t$ are highly unlikely to be identified as robust after further preference elicitation. Therefore, the open interval $\left(\underline{V}_{i'}^{\#i}, \bar{V}_{i'}^{\#i}\right)$ is used for $g_A(i)$ in the MCDA case so that options $t$ are counted as practically dominated.

[4] The table below provides a numerical illustration for the example depicted in Fig. 2. Three participants assess four options on two criteria and the median is used to predict the group's agreement. The borderline takes the value $\hat{\varphi} = \max\{V_i^*\} = 66$. Furthermore, as an additional notation, let $\bar{v}_{i'j}^{\#i}$ be the upper bound of $v_{i'j}$ when assuming that $v_{ij}$ takes the predicted value $v_{ij}^*$.

|  | *Criterion 1 ($w_1 = 0.5$)* | | | *Criterion 2 ($w_2 = 0.5$)* | | | | | |
| --- | --- | --- | --- | --- | --- | --- | --- | --- | --- |
|  | *Participant 1* | *Participant 2* | *Participant 3* | *Participant 1* | *Participant 2* | *Participant 3* | $\underline{V}_i$ | $\bar{V}_i$ | $V_i^*$ |
| Option 1 | 6 | 94 | 68 | 64 | 100 | 50 | 28 | 97 | 66 |
| Option 2 | 0 | 10 | 90 | 0 | 36 | 62 | 0 | 76 | 23 |
| Option 3 | 43 | 12 | 64 | 50 | 4 | 47 | 8 | 57 | 45 |
| Option 4 | 38 | 78 | 84 | 36 | 38 | 64 | 37 | 74 | 58 |

It is shown in the following that the measure $g_A$ for the Value Agreement of option 4 takes the value of 1. Option 4 dominates option 2 on criterion 2 because all three participants $k$ assigned a higher score to $v_{42k}$ than to $v_{22k}$. Using the median, the predicted score for option 4 on criterion 2 is $v_{42}^* = 38$, which becomes the upper bound for option 2 on criterion 2; i.e. $\bar{v}_{22}^{\#4} = 38$. $\bar{v}_{21}^{\#4} = \bar{v}_{21} = 90$ because option 4 does not dominate option 2 on criterion 1. It follows that the upper bound for option 2, when assuming the predicted agreement for option 4, is $\bar{V}_2^{\#4} = w_1 \bar{v}_{21}^{\#4} + w_2 \bar{v}_{22}^{\#4} = 64$, which is smaller than the borderline $\hat{\varphi}$. The upper bound of option 3 $\bar{V}_3$ is already smaller than $\hat{\varphi}$, and $\bar{V}_4^{\#4} = V_4^* = 58$ as well. There are no dominance relationships between option 1 and 4, which means $\bar{V}_1^{\#4} = \bar{V}_1 = 97$. It follows that only option 1 intersects the borderline when assuming that the group agrees on the predicted scores $v_{ij}^*$ for option 4, meaning that $g_A(4) = 1$.

[5] When testing Liesiö et al.'s PDA preference programme (2007) for incomplete weight information, we noticed that their algorithm does not identify all dominated portfolios and therewith all dominated options. For instance, it is possible that a portfolio $p_1$ is dominated by portfolio $p_2$ or $p_3$ for all feasible weight combinations, but there is always a feasible weight combination such that $p_1$ is not dominated by $p_2$ and a feasible weight combination such that $p_1$ is not dominated by $p_3$. As Liesiö et al.'s algorithm only uses pairwise comparisons of portfolios, it does not recognise that $p_1$ is actually dominated. To test whether a portfolio $p_1$ is dominated by any other set of portfolios for all feasible weight combinations, one can solve a linear programme with arbitrary objective function where (i) the weights are constrained by their upper and lower bounds and (ii) there is a linear constraint for each other's portfolio such that $p_1$ is not dominated by this portfolio. If the linear programme has no feasible solution for the weights, there is no feasible weight allocation so that $p_1$ is not dominated—hence $p_1$ is dominated.

[6] For the PDA case, the preference programme from Liesiö et al. (2007) is given a time limit of 2 seconds. The algorithm is written in VB.NET x64 release and ran on a 2.8 GHz processor with 4GB RAM.

[7] As $c_i = 1$ for all candidates $i \in I$, the value-for-money ratio bars depicted in Fig. 5 are identical with the candidates' value score bars $[\underline{V}_i, \bar{V}_i]$.

[8] Given the limited information about a candidate's personality fit before the interview, the panel members agreed to be generous when appraising only the candidates' written applications and scored most participants between 70 and 90 on this criterion in step 2 of the recruitment process.